\documentclass[conference]{IEEEtran}
\IEEEoverridecommandlockouts
\usepackage{hyperref}
\hypersetup{
    colorlinks=true,
    linkcolor=blue,
    filecolor=magenta,      
    urlcolor=cyan,
    pdftitle={Overleaf Example},
    pdfpagemode=FullScreen,
    }

\usepackage{cite}

\usepackage[official]{eurosym}
\usepackage{amsthm}

\usepackage{amsmath,amssymb,amsfonts}
\usepackage{graphicx}

\usepackage{textcomp}
\usepackage{xcolor}
\def\BibTeX{{\rm B\kern-.05em{\sc i\kern-.025em b}\kern-.08em
    T\kern-.1667em\lower.7ex\hbox{E}\kern-.125emX}}

\usepackage[normalem]{ulem}

\usepackage[nolist]{acronym} 

\usepackage{tcolorbox}

\usepackage{framed,color}
\definecolor{shadecolor}{rgb}{1,0.8,0.3}
    
\newcommand{\fref}[1]{Fig.~\ref{#1}}
\newcommand{\sref}[1]{Section~\ref{#1}}
\newcommand{\pref}[1]{Section~\ref{#1}}
\newcommand{\tref}[1]{Table~\ref{#1}}

\begin{document}

\begin{acronym}[HBCI]

\acro{ML}{Machine Learning}
\acro{KPI}{Key Performance Indicator}
\acro{MQTT}{Message Queuing Telemetry Transport}
\acro{SLAM}{Simultaneous Localization and Mapping}
\acro{M2M}{Machine-to-Machine}
\acro{AI}{Artificial Intelligence}
\acro{DT}{Digital Twin}
\acro{XR}{Extended Reality}
\acro{IRF}{Intelligent Radio Fabric}
\acro{ng-DT}{Next-generation Digital Twin} 
\acro{O-RAN}{Open Radio Access Networks}
\acro{LLMs}{Large Language Models}
\acro{ML}{Machine Learning}
\acro{CE}{Channel Estimation}
\acro{AI}{Artificial Intelligence}
\acro{RAN}{Radio Access Network}
\acro{MIMO}{Multiple Input, Multiple Output}
\acro{SDR}{Software-Defined Radio}
\acro{OFDM}{Orthogonal Frequency-Division Multiplexing}
\acro{mmWave}{Millimeter Wave}
\acro{cmWave}{Centimeter Wave}
\acro{THz}{Terahertz}
\acro{PHY}{Physical Layer}
\acro{CSI}{Channel State Information}
\acro{UE}{User Equipment}
\acro{CSI-RS}{Channel State Information-Reference Signal}
\acro{CQI}{Channel Quality Indicator}
\acro{PMI}{Precoding Matrix Indicator}
\acro{RI}{Rank Indicator}
\acro{SE}{Spectral Efficiency}
\acro{DFT}{Discrete Fourier Transform}
\acro{SVD}{Singular Value Decomposition}
\acro{FC}{Fully Connected}
\acro{WMMSE}{Weighted Minimum Mean Squared Error}
\acro{SGCS}{Sum-Rate Gain Comparison Study}
\acro{NMSE}{Normalized Mean Squared Error}
\acro{PC}{Personal Computer}
\acro{RSRP}{Reference Signal Received Power}
\acro{ADC}{Analog to Digital Converter}
\acro{SDR}{Software Defined Radio}
\acro{RF}{Radio Frequency}
\acro{SISO}{Single Input Single Output}
\acro{BS}{Base Station}
\acro{RSS}{received signal strength}
\acro{ULA}{Uniform Linear Array}
\acro{UPA}{Uniform Planar Array}
\acro{FFT}{Fast Fourier Transform}
\acro{IFFT}{Inverse Fast Fourier Transform}
\acro{CP}{Cyclic Prefix}
\acro{AWGN}{Additive White Gaussian Noise}
\acro{TDD}{time division duplex}
\acro{GRC}{GNU Radio Companion}
\acro{AOD}{Angle of Departure}
\acro{AoA}{Angle of Arrival}
\acro{LOS}{Line of Sight}
\acro{NLOS}{Non-Line of Sight}
\acro{DQN}{Deep Q-Network}
\acro{ReLU}{Rectified Linear Unit}
\acro{PDF}{Probability Density Function}
\acro{OOT}{Out of Tree}
\acro{NN}{Neural Network}
\acro{DNN}{Deep Neural Network}
\acro{CNN}{Convolutional Neural Network}
\acro{GNN}{Graph Neural Network}
\acro{NN}{Neural Network}
\acro{ML}{Machine Learning}
\acro{ZF}{Zero Forcing}
\acro{FLOPS}{Floating Point Operation per Second}
\acro{MSE}{Mean Squared Error}
\acro{BER}{Bit Error Rate}
\acro{DL}{Downlink}
\acro{UL}{Uplink}
\acro{TX}{Transmitter}
\acro{RX}{Receiver}
\acro{LS}{Least Square}
\acro{MMSE}{Minimum Mean Squared Error}
\acro{FPGA}{Field Programmable Gate Array}
\acro{SNR}{Signal to Noise Ratio}
\acro{SINR}{Signal to Interference plus Noise Ratio}
\acro{MISO}{Multiple-Input Single-Output}
\acro{FDM}{Frequency Domain Multiplexing}
\acro{IF}{Intermediate Frequency}
\acro{1D}{one-dimensional}
\acro{LMMSE}{Linear Minimum Mean Squared Error}
\acro{RNN}{Recurrent Neural Network}
\acro{SLL}{Side-Lobe Level}
\acro{LLM}{Large Language Model}
\acro{IIoT}{Industrial Internet of Things}
\acro{IoT}{Internet of Things}
\acro{XR}{Extended Reality}
\acro{GPU}{Graphical Processing Unit}
\acro{PDSCH}{Physical Downlink Shared Channel}
\acro{LAN}{Local Area Network}
\acro{FC}{Fully Connected}
\acro{lwm}[LWM]{Large Wireless Model}
\acro{MWC}{Mobile World Congress}
\acro{RLS}{Recursive Least Squares}
\acro{PDSCH}{Physical Downlink Shared Channel}
\acro{MoE}{Mixture of Experts}
\acro{MCS}{Modulation and Coding Scheme}
\acro{AMC}{Adaptive Modulation and Coding}
\acro{PAAM}{Phased-Array Antenna Module}
\acro{MU}{Multi-User}
\acro{DRL}{Deep Reinforcement Learning}
\acro{HW}{Hardware}
\acro{SW}{Software}
\acro{URLLC}{Ultra Reliable Low Latency Communications}
\acro{HARQ}{Hybrid Automatic Repeat Request}
\acro{WF}{Water Filling}
\acro{NR}{New Radio}
\acro{gNB}{next-generation Node B}
\acro{MWh}{Megawatt hour}
\acro{kWh}{Killowatt hour}
\acro{QoS}{Quality of Service}
\acro{FNN}{Feedforward Neural Network}
\acro{FCN}{Fully-Connected Network}
\acro{DPA}{Dynamic Power Allocation}
\acro{TF}{Transformer}
\acro{RE}{Resource Element}
\acro{RNN}{Recurrent Neural Network}
\acro{ViT}{Vision Transformer}
\acro{AMP}{Automatic Mixed Precision}
\acro{DDPG}{Deep Deterministic Policy Gradient}
\acro{RAM}{Random Access Memory}
\acro{CPU}{Central Processing Unit}
\acro{API}{Application Programmable Interface}
\acro{FLS}{Fast-learning System}
\acro{SLS}{Slow-learning System}

\end{acronym}

\title{AI-Programmable Wireless Connectivity: Challenges and Research Directions Toward Interactive and Immersive Industry
}


\author{
\IEEEauthorblockN{Haris Gacanin, 
\IEEEmembership{Fellow, IEEE}\\
RWTH Aachen University}
} 

\maketitle
\thispagestyle{plain}
\pagestyle{plain}

\begin{abstract}
This vision paper addresses the research challenges of integrating traditional signal processing with \ac{AI} to enable energy-efficient, programmable, and scalable wireless connectivity infrastructures. While prior studies have primarily focused on high-level concepts, such as the potential role of \ac{LLM} in 6G systems, this work advances the discussion by emphasizing integration challenges and research opportunities at the system level. Specifically, this paper examine the role of compact AI models, including Tiny and Real-time Machine Learning (ML), in enhancing wireless connectivity while adhering to strict constraints on computing resources, adaptability, and reliability. Application examples are provided to illustrate practical considerations, highlighting how AI-driven signal processing can support next-generation wireless networks. By combining classical signal processing with lightweight AI methods, this paper outlines a pathway toward efficient and adaptive connectivity solutions for 6G and beyond.

\end{abstract}

\begin{IEEEkeywords}
Intelligent radio fabric, XR, Industry 5.0, AI-Native wireless, TinyML, sensing, distributed AI.
\end{IEEEkeywords}

\section{Introduction}\label{sec:intro}
Recently, it has been widely discussed that \ac{AI} applications will play a central role in driving the operations and management of 6G wireless connectivity \cite{wwf}. Different industries (e.g., production, healthcare, transportation) have started to accelerate the adoption of digital data technologies, such as \ac{DT} and \ac{AI}, to enable ubiquitous human-machine interactions. However, with the development of mobile \ac{AI} applications, \ac{AI} computing and dynamic sensing by using cyber-physical systems, further technological evolution is expected toward immersive and interactive human presence \cite{nokia2024ai, ericsson-ainative}. This is envisioned by the integration of haptics, extended reality, and/or a holographic computing system in 6G connectivity. 

\begin{figure}
    \centering    \includegraphics[width=\linewidth]{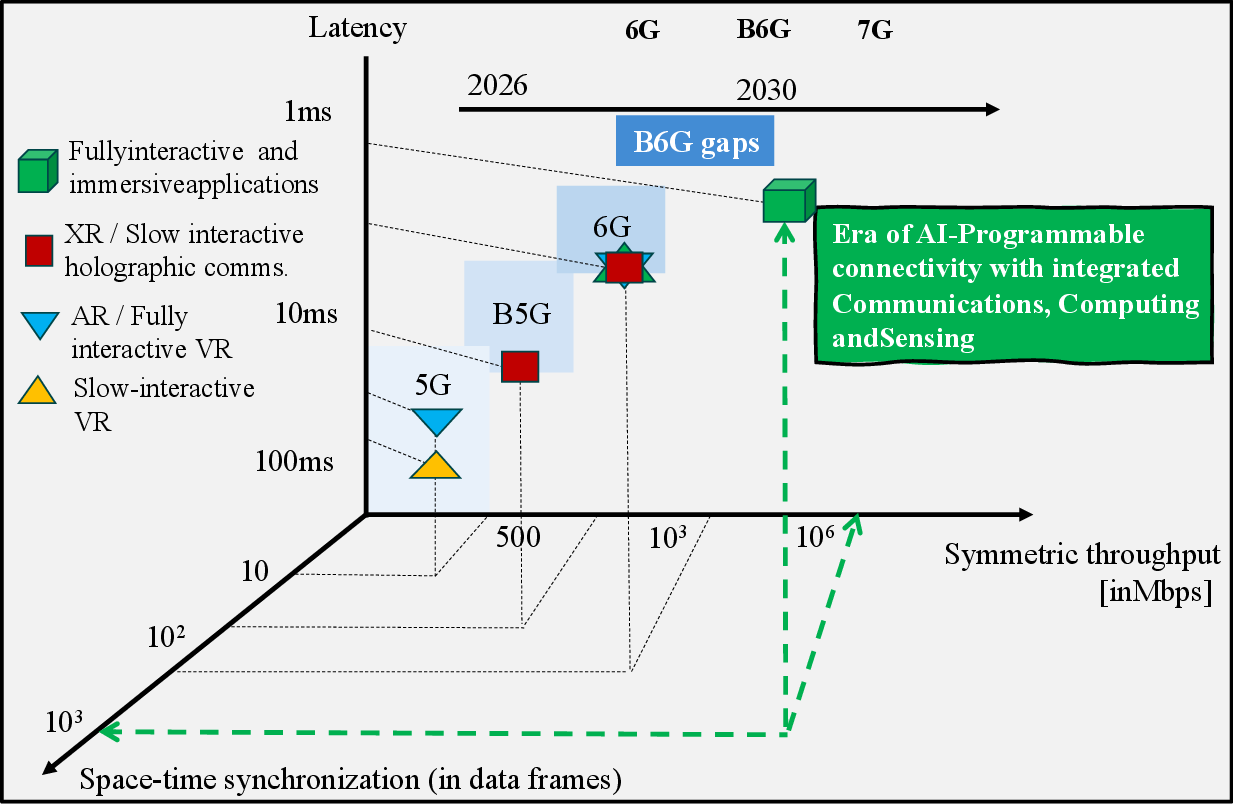}
    \caption{6G performance requirements for interactive mobile \ac{AI} applications.}
    \label{fig:requirements}
\end{figure}

To enable a realistic perception of human presence, it is necessary to transmit three-dimensional images with audio or other modalities to excite the human senses by using extreme communication-performance and computing-demanding devices. For mobile \ac{AI} applications, a fully immersive room's visual data can require up to 8.7 Tbps data streams. The compression retina resolution VR-headsets targeting the human field-of-view can reduce the requirement to 60 Gbps. Moreover, the pupil tracking and compression further reduce this down to 3.3 Gbps -- however, these are still remarkably high connectivity requirements \cite{holocomms}. Besides extreme throughput challenged by multi-band propagation (e.g., cm/mm/THz) and ultra-low latency (i.e., $\leq$1 ms) requirements, the high communications-resource-efficiency with scalable (i.e., 50B+ devices) and energy-efficient computing for distributed learning (e.g., \ac{LLM}, \ac{DRL}) needs to be considered \cite{wwf, ai4coms}. While traditional signal processing methods are, in theory, performance-efficient, they lack the adaptability to handle dynamic, data-rich environments. Conversely, standalone \ac{AI} computational methods often overlook the communication constraints of wireless applications, leading to suboptimal energy use and transmission errors. 

This paper lists the key wireless system requirements (latency, throughput, synchronization) in \fref{fig:requirements}, which are necessary for interactive and mobile \ac{AI} supported applications from the ultra-dense connectivity design perspective. In principle, a distribution of a mobile \ac{AI} model across devices introduces challenges related to synchronization, uplink/downlink communication, and latency \cite{tinyML1}. Consequently, spatial computing (i.e., \ac{XR}) supported with mobile \ac{AI} is revolutionizing wireless system management by fusing physical and digital domains for enhanced visibility, analysis, and optimization. Thus, supporting interactive and mobile \ac{AI} applications will require wireless connectivity with an adaptive bandwidth for communications over extremely reliable wireless links, while integrating distributed \ac{AI} computing and sensing over non-homogeneous devices.

\subsection{An example of future mobile \ac{AI} application}
Let's consider the following example. Imagine that your vital functions are monitored by your home network, and that it senses an unusual situation before you even feel sick. The network deduces your ``state'' and initiates an interactive in-home hospital application, while transforming the digital wall's content into an in-home emergency room. The network collects data and provides \ac{AI} diagnostics, and a holographic doctor service is activated. In the remote hospital, the (real) doctor sees your holographic body and personal data. A remote interactive examination starts. The diagnosis is set, and the medicine requested from a local pharmacy is delivered to your home. In some situations, a medical team in an autonomous emergency vehicle is dispatched for hospitalization in a local health center. The network driven by your senses initiated a virtual hospital application and responded to your needs, providing the health experts a larger window of response possibilities. 

In such a future, wireless connectivity is expected to face a massive increase in heterogeneity of data generated by non-homogeneous devices (i.e., devices with device storage and \ac{CPU}/\ac{GPU} computing capabilities) and mobile \ac{AI} applications, where demands on communications and computing resources are transient, unpredictable, and dynamic. 
Unlike the big-data paradigm, where we place a massive amount of data in the cloud, the application in our example relies on distributed connectivity and sensing supported by on-device and edge-cloud computing to process and facilitate different data flows, while simultaneously, numerous connectivity nodes and application devices can share their data, computing, and storage resources. 

\begin{figure}[t]
\centerline{\includegraphics[width=\linewidth]{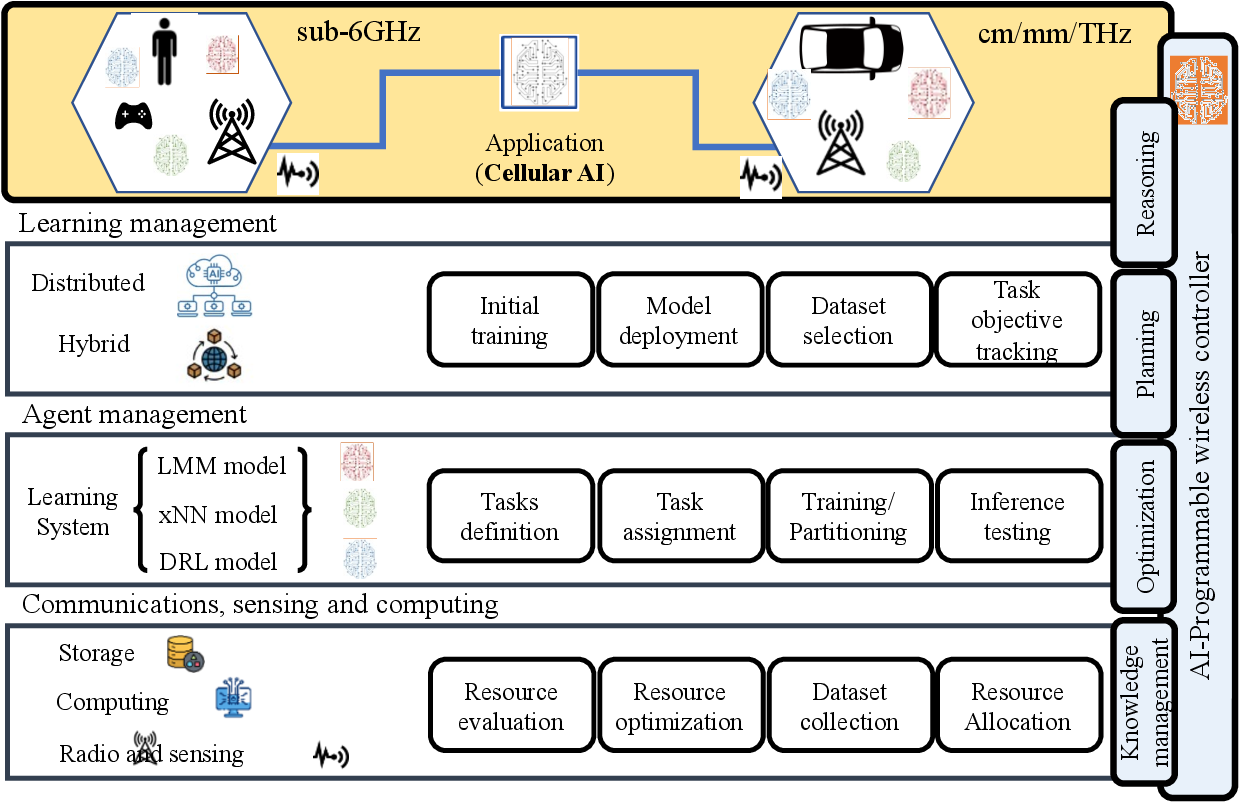}}
 \caption{Intelligent radio fabric for connected intelligence as an interactive application enabler.}
 \label{fig:irf}
 \end{figure}

\subsection{Technology vision}\label{sec:vision}
To unfold such applications, this paper envisions wireless connectivity beyond 2030 as an \ac{IRF} -- a distributed and \ac{AI}-programmable communication and computing infrastructure that harvests the computational power of a continuously growing number of devices. This paper illustrates high-level functions of the \ac{IRF} in \fref{fig:irf}. The utilization of device storage (e.g., \ac{RAM}, \ac{FPGA}), computing (e.g., \ac{CPU}, \ac{GPU}), and radio (e.g., subcarrier, power) resources is controlled through IRF's \ac{AI}-Programmable controller, which orchestrates the Learning and Agent management functions. The controller orchestrates the distributed computing power of connected devices to realize \ac{IRF} vision. This paper designs distributed learning by integrating embedded learning on devices (i.e., low-power \ac{ML} such as Tiny \ac{ML}), Real-time \ac{ML}, and edge-based learning/computing.

The paper provides the data measurements through wireless connectivity (i.e., communications, sensing, and computing functions) to the Agent management layer, where different agents (i.e., LLM, TinyML, DRL \cite{tinyML2}) are designed to compute their models in a distributed or hybrid architecture coordinated by the Learning management layer. The Learning management layer decides if the distributed computation (training and inference) of \ac{AI} models is performed locally on devices (i.e., Agentic AI \cite{agenticAI}) or if it partially transfers the data to the edge computing in a hybrid learning architecture. Choosing between hybrid and distributed learning architectures depends on the computing and storage capabilities of the connected devices, service quality, and orchestration capability of the 6G system \cite{3gpp-ai}. For example, an ultra-low latency requirement of below 1 ms is a major requirement not only as an enabler of mobile \ac{AI} applications, but also the communications between \ac{AI}-enabled devices \cite{6g-ai-standard}. 

\begin{table}[t!]
  \begin{center}
    \caption{Examples of Inputs/Outputs/Timing interfaces for the \ac{IRF} functions.}
    \label{tab:api}
    \begin{tabular}{|p{1.7cm}|p{2.2cm}|p{2cm}|p{1.5cm}|}
      \hline
      \textbf{Function} & \textbf{Inputs (to Edge)} & \textbf{Outputs (to Device)}& \textbf{Expected latency}\\ \hline
      PHY/MAC closed-loop adaptation & Channel quality and state indicators, received signal level, buffer status, traffic pattern & Modulation and coding, beam index, power control & 0.125 - 1 ms (sub slot-level) \\ \hline
      RF/environment sensing & Range-Doppler information, angle profile, device position & Sensing task config (waveform, bandwidth) & 1 - 10 ms \\ \hline
      Distributed inference with split learning & Local feature vectors, model gradients, device energy state & Model partitioning decision, quantization level, and offloading & 1 - 50 ms \\ \hline
      Cross-layer optimization & Mobility state, battery level, application demand & QoS profile, wake/sleep schedule & 10 - 100 ms \\ \hline
      Space-Time synchronization & Local timestamp, position, clock drift & Global time reference update & $<$1 µs jitter \\ \hline
    \end{tabular}
  \end{center}
\end{table}

  \begin{figure*}[t]
\centerline{\includegraphics[width=5in]{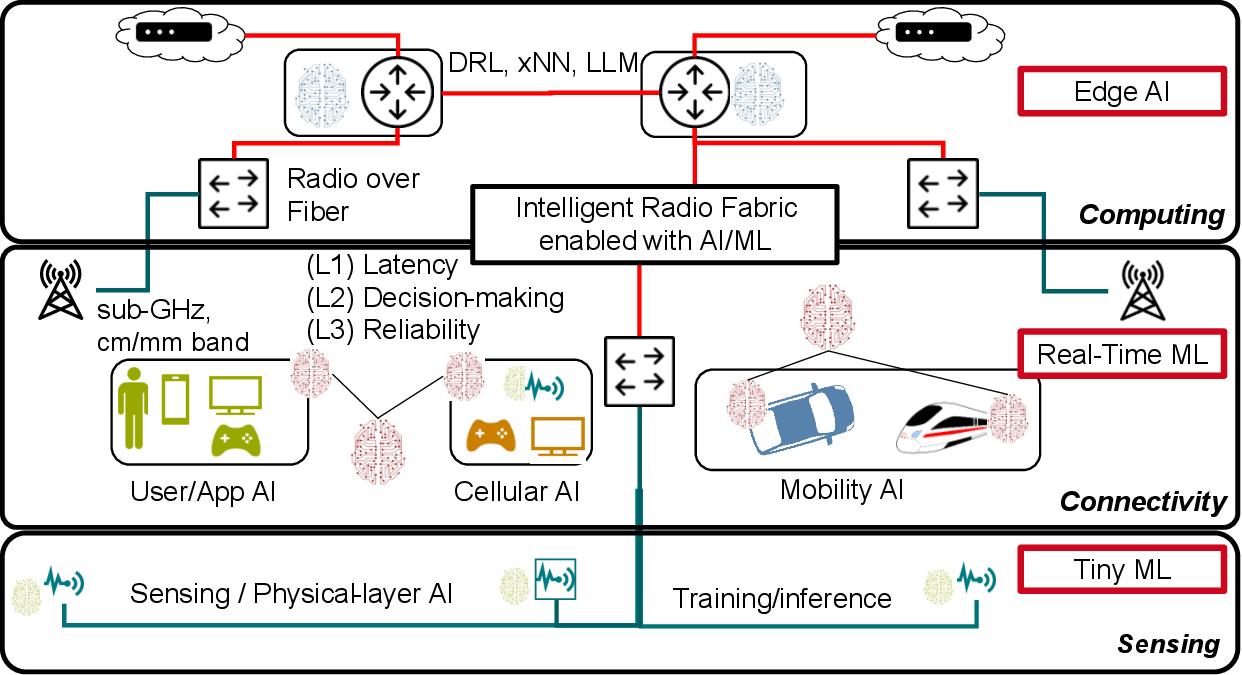}}
 \caption{Distributed \ac{AI} across non-homogeneous devices.}
 \label{fig:connectedAI}
 \end{figure*}
 
In \fref{fig:connectedAI}, the components of distributed \ac{AI} functions are illustrated across non-homogeneous devices, while \tref{tab:api} illustrates the expected network \ac{API}s to for \ac{AI}-enabled devices \cite{6g-ai-standard}. Here, the Real-time \ac{ML} automates the efficient communications with low overhead/latency, and computing workloads in Agent and Learning management for \ac{AI} inference and training. The workload is controlled via the connectivity function, splitting a model execution across connectivity and edge, while utilizing different communications, computing, and sensing capabilities in the vicinity of mobile \ac{AI} devices. 


\subsection{Goals of this paper}
This vision paper discusses research challenges of integrating traditional signal processing with \ac{AI} computational methods to create energy-efficient, programmable, and scalable wireless connectivity infrastructure. Recently, in \cite{wwf}, a quantitative characterization of the 6G applications of \ac{AI} was discussed with a focus on high-level concepts of \ac{LLM} integration with connectivity infrastructure. However, this paper, goes a step further and discusses integration challenges and potential research directions with application examples. Consequently, wireless connectivity with integrated \ac{AI} is defined by using ``small'' models such as Tiny/Realtime ML, while considering system-level limitations (e.g., computing, adaptability, and reliability). Finally, the goals of this paper are summarized as follows: 
\begin{itemize}
\item First, the technological challenges of \ac{IRF} are discussed in \sref{sec:challenges}, while focusing on practical implications of computing, adaptability, and reliability challenges on mobile \ac{AI} applications. 
\item Enabling technologies are discussed for integrating distributed
and AI-programmable communications, computing, and sensing in \sref{sec:tech}, where embedded on-device and edge-based learning supports mobile \ac{AI} applications. 
\item Finally, this paper conceptualize an implementation example of integrated wireless connectivity with a spatial computing system in \sref{sec:dev}.
\end{itemize}

\section{Technology Challenges of Mobile \ac{AI} Applications}\label{sec:challenges}
\begin{figure}[t]
\centerline{\includegraphics[width=\linewidth]{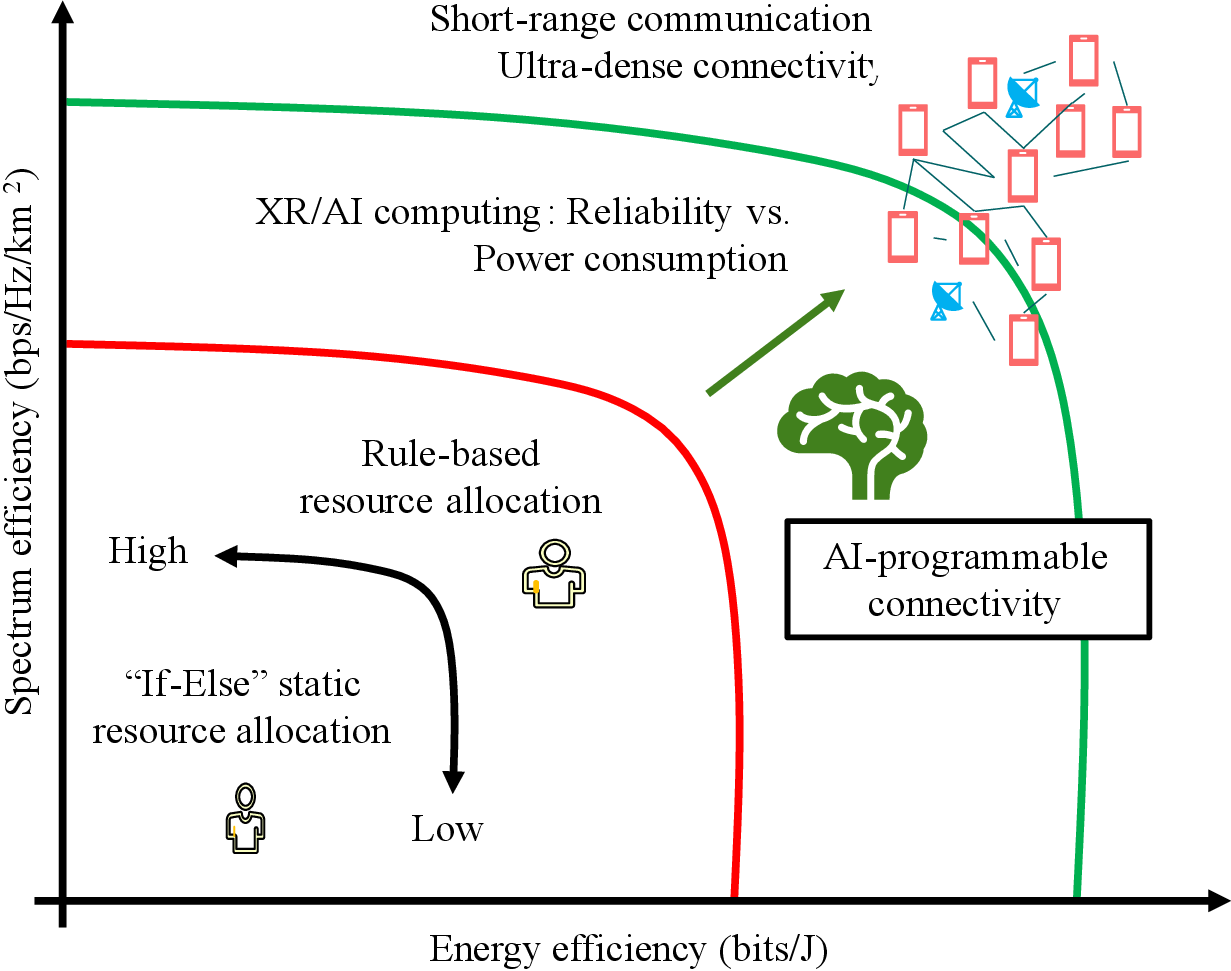}}
\caption{Trade-off between spectrum and energy efficiency leading to AI-based short-range connectivity.}
\label{fig:tradeoff}
\end{figure}

%
Following the above-discussed requirements, for the design of a real-time decision-making in latency-constrained interactive and mobile \ac{AI} applications, the core three limitations of the AI-programmable wireless connectivity are defined: 
\begin{description}
    \item[\textbf{(L1)}] \textbf{Computing}: Reliance on energy-intensive GPUs for AI/ML computing, which introduces critical end-to-end latency above 0.1 ms (e.g., Open RAN real-time processing 0.1$\sim$1 ms); 
    \item[\textbf{(L2)}] \textbf{Adaptability}: Sub-$\mu$s decision-making for ultra-dense and distributed learning in wireless systems (e.g., 4.17 $\mu$s OFDM symbol timing); and
    \item[\textbf{(L3)}] \textbf{Reliability and Coverage}: Consistent and accurate delivery of Gbps symmetric (downlink/uplink) throughput in distributed wireless systems.
\end{description}
Now, under these limitations, the question we pose is, can we simultaneously improve both spectrum and energy efficiency while maintaining coverage satisfactorily? In \fref{fig:tradeoff}, we illustrate a trade-off relationship between the increase of spectral efficiency (SE) with densification utilizing multi-band large-scale antenna systems (in bps/Hz/km$^2$) and energy-efficiency (EE) with energy-constraint modulation and resource allocation (in bits/J), while keeping the transmit power as a parameter.

In ultra-dense wireless networks, SE and EE are tightly coupled, with higher densification or transmit power improving SE while supporting 1–10 Gbps per user and 10–30 bit/s/Hz per cell in networks exceeding 100 small cells/km$^2$, but increasing energy consumption due to interference management and active radios. EE targets range from 1–10 $\mu$J/bit for uplink/downlink, with microcells consuming 10–50 W and pico/femtocells 1–5 W, while edge computing adds 5–500 W depending on device. Adaptive strategies such as beamforming, dynamic offloading, and sleep modes can maintain $>70\%$ of peak SE while achieving $>80\%$ of peak EE, balancing throughput and energy in immersive XR applications.

For immersive XR applications, the higher throughput and lower latency budget also invoke higher energy consumption and more network resources. For example, the more efficient XR computing (e.g., on-device) can lower data throughput requirements, but requires more energy and higher reliability. Therefore, a trade-off between the network performance and on-device computing needs to be considered.

\subsection{Robust and adaptive communication in real time}\label{sec:responsivenessChallenge}
The ability of 6G wireless to respond to rapidly changing conditions is the capability of radio interfaces to dynamically (re-)optimize based on environment percepts, overcoming the limitations of both (\textbf{L1}) and (\textbf{L2}) \cite{ai4coms}. For example, the energy-constrained devices (i.e., wearables, XR headsets, and smart sensors) need to engage in context-aware communication and decision-making without compromising battery life or reliability. Thus, a critical connectivity requirement is the support of ultra-low-power adaptive communications, which must account for a wide range of devices.

Moreover, adaptiveness with distributed systems having non-homogeneous devices will require robust decision-making at a microsecond scale (i.e., \ac{OFDM} symbol duration from 66 down to 1 $\mu$s), with radio latencies approaching the millisecond scale \cite{wwf, 3gpp-ai, 3gpp-xr}. These stringent demands necessitate the development of open, standardized interfaces that can support tight integration with AI-programmable (i.e., AI-Native) control systems, including those driven by cyber-physical feedback (e.g., vision, gesture, or motion-based input) \cite{6g-ai-standard}.

\subsection{Training in Real Time}\label{sec:learningChallenges}
The mobile \ac{AI} in contemporary wireless applications mostly rely on centralized learning models \cite{wwf}. While suitable for high-level network monitoring and offline optimization, the centralized computing falls short in the context of real-time control and adaptation in dynamic and latency-constrained scenarios. For instance, in our example, the optimal configuration of the distributed connectivity may shift dramatically with users' mobility, evolution of sensor readings, and setup of new wireless links. The shift to distributed learning faces challenges with a low-cost multiple-objectives optimization in terms of throughput, computing, and data dependencies over non-homogeneous devices. Consequently, one needs to address the distribution of learning tasks across a diverse set of devices, while keeping the communications constraints in \sref{sec:communicationEfficiency}. While maintaining convergence and model quality driven by the immersive application requirement, the training tasks must be partitioned and scheduled across devices with varying capabilities. At the same time, inference tasks often required in real-time and on-device learning must be adapted to the local computational and energy constraints in \sref{sec:computingChallenges}. Here, the integration of Tiny \ac{ML} and Real-time \ac{ML} with 6G connectivity in \fref{fig:connectedAI} is an open research topic, where the tradeoffs between energy efficiency and inference/training accuracy are not yet understood. On-device training/learning in real-time needs to be supported by \ac{AI}-programmable data collection, where data is collected and acted upon at the right location and time, without overprovisioning or creating new bottlenecks elsewhere in the network (see discussion in \sref{sec:dataChallenges}).

\begin{figure}[t]
\centerline{
\includegraphics[width=3.5in]{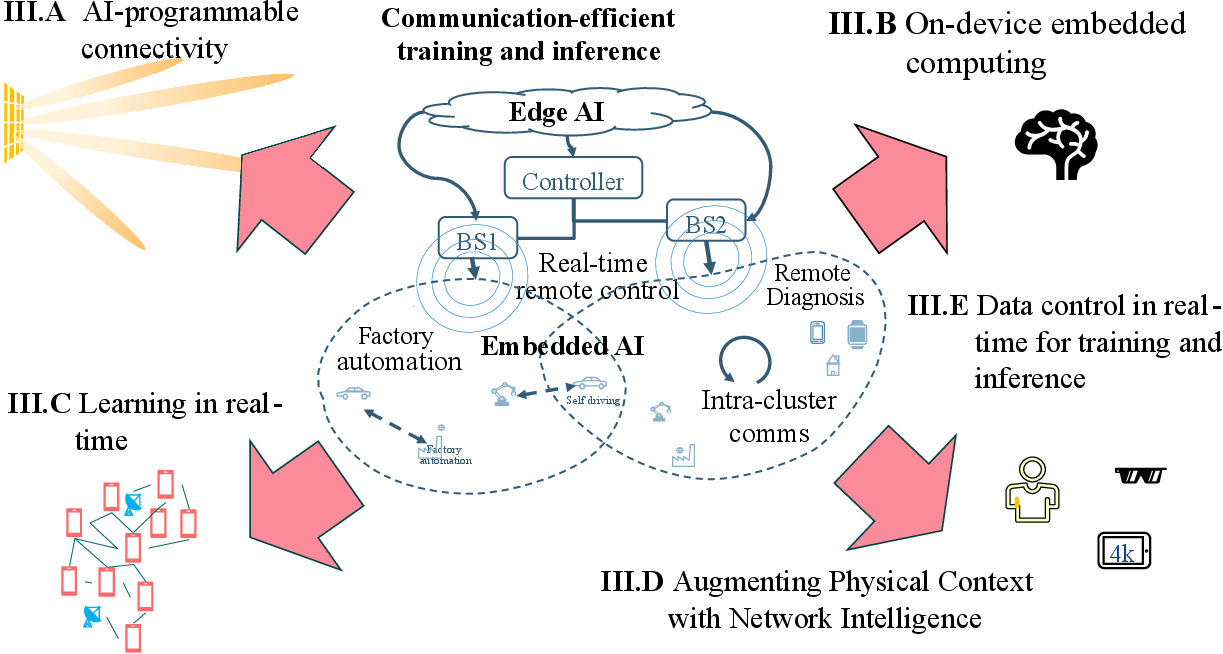}
}
\caption{Technology components for building IRF and related research directions.}
\label{fig:directions}
\end{figure}

\subsection{Device Communications Efficiency}\label{sec:communicationEfficiency}
Enabling large-scale \ac{AI}-programmable distributed connectivity requires efficient device-to-device (i.e., short-range) communications. 

The first challenge is the design of a communication-efficient protocol for distributed model training across non-homogeneous devices, while coping with the unreliability and non-stationarity of wireless channels  \cite{ericsson-ainative,nokia2024ai}. Ideally, resource allocation algorithms for distributed computing and storage should be utilized to enable on-device learning and training. However, the high-speed, low-latency, and high-precision synchronized communications between devices becomes a bottleneck as discussed in \fref{fig:requirements}. To mitigate this, distributed or federated learning can be used to collaboratively train models without exchanging raw data and reducing uplink communication overhead \cite{wwf}. Nonetheless, federated learning introduces new constraints related to synchronization of devices' communications, leading to problems with model convergence across non-independent and identically distributed data sources.

In parallel to training, the second challenge is the design of communication-efficient distributed inference. The execution of inference tasks needs to be communicated with stringent latency, synchronization, and reliability across connected devices and edge, as illustrated in \fref{fig:connectedAI}. For example, \ac{AI} models can be partially executed on the devices to extract features, while it is not clear how to design an efficient communications protocol with optimal accuracy, without the need to transmit to edge servers for final decision-making. Therefore, the non-homogeneous and resource-constrained nature of connected devices challenges communications of ``optimal'' definition, assignment/partitioning, training, and real-time inference testing in \fref{fig:irf}. For instance, this includes minimizing redundant data transmission and spatiotemporal computation supported by edge computing, posing non-trivial trade-offs between model accuracy, responsiveness, and energy efficiency.

Our application example requires support of {adaptive packet and short-range connectivity}, essentially meeting the stringent latency and reliability limitation of such critical applications in \textbf{(L2)} and \textbf{(L3)}. However, achieving reliable short-range connectivity goes beyond maintaining high signal strength. A major challenge lies in the uncoordinated spatiotemporal distribution of end devices and the often sub-optimal placement of connectivity nodes (e.g., 3GPP or IEEE 802.11 access points). 

On the other hand, ultra-densification introduces interference (see \fref{fig:tradeoff}), which leads to intermittent connectivity, due to massive overlapping of the cells \cite{self-opt}. In mobile \ac{AI} applications, even a short communication outage can severely disrupt coordination between sensing devices, robotic actuators, and connectivity devices \cite{wwf}. For example, this is particularly critical for time-sensitive and interactive applications like collaborative robotic rescue, which rely on continuous, low-latency data exchange to operate safely and effectively. To enable the function where adaptive packet sizes will be used to split between communications and sensing phases, one needs to consider adaptive bandwidth management. This leads to challenges related to the tradeoff between energy efficiency  (i.e., computing split) and spectral efficiency (i.e., bandwidth management).

\subsection{Computing efficiency}\label{sec:computingChallenges}
\ac{O-RAN} architecture relies on (near-)homogeneous hardware and computing devices, integrating general-purpose and dedicated devices for a range of functions such as resource management, \ac{AI}, remote control, etc. Let's consider the centralized learning model where \ac{GPU} (e.g., H100 GPU with a peak power consumption of 700 W) is annually utilized 60\%. The estimated electricity consumption is 3,68 \ac{MWh} per year, which is equivalent to the annual consumption of the average German household. Assuming that 3,5 Million \ac{GPU}s are deployed by late 2025, they can consume 12,880,000 \ac{MWh} of electricity per year (ca. \euro{3,456} Billion for the average price of 26 cents per kWh).

This paper questions whether centralized learning models are feasible for future wireless applications, where decision-making with non-homogeneous devices in real time is needed, or instead, one must consider distributed computing supporting learning based on frequently arriving, small-volume data streams with minimal latency, as discussed in \sref{sec:learningChallenges}. Since these spatially distributed devices differ significantly in communications capabilities, computing and memory availability, mobility patterns, and energy limits, the allocation of computing and storage resources presents a major barrier to real time energy-efficient of \ac{AI} models in 6G wireless \cite{tinyML3}.

\section{Research Directions for Realization of Intelligent Radio Fabric}\label{sec:tech}
Next, the major research directions are discussed necessary to address the above challenges.

\subsection{AI-programmable connectivity}
In our virtual emergency room example, we require short-range connectivity where the environment knowledge is updated per locally ``selected'' communication devices. For example, the spatiotemporal ``communication- and computing-diversity'' can be studied to advance \ac{AI} distributed resource allocation techniques. Because of the increased interference (see \pref{sec:communicationEfficiency}) due to densification of inter- and intra-communications clusters in \fref{fig:directions}, an ultra-high reliability needs to be maintained by distributed multi-user \ac{MIMO} beamforming, and near-field \ac{MIMO} processing. This is because, due to communication errors, distributed systems are often more difficult to develop, control, and manage in comparison with centralized systems. For example, in ultra-dense deployments to increase the transmission reliability and spectrum efficiency, besides efficient deployment, the communication between devices over long distances or with physical obstacles needs to be avoided. Thus, the goal of AI-programmable short-range connectivity leads to \ac{IRF} that is self-constructible through distributed computing and learning. Therefore, the programmability can be achieved with \ac{AI}-native standardization and ``openness'' of the radio access for AI-enabled interfaces \cite{6g-ai-standard}. Finally, addressing the gap of autonomous deployment planning, the programmable connectivity needs to address automated (re)deployment strategies.

\subsection{On-device embedded computing}\label{sec:embeddedComputing}
Instead of demanding cloud computing facilities, a learning framework is defined to continuously adapt connectivity using distributed devices as sensors. Specifically, all connected devices are computing nodes with varying capabilities, with data processing shared between the cloud and the devices. \ac{IRF} facilitates local computing and storage at different locations in parallel, where and when needed -- namely, embedded on-device learning. 

To support 6G applications, hybrid computing combining Tiny ML methods with classical signal processing, i.e., compressed sensing for efficient channel estimation and subspace tracking for dynamic adaptation, can be used to reach real-time decision-making while improving energy efficiency. Together, Tiny ML and real-time ML transform 6G from a high-speed transport layer into a context-aware, distributed intelligence fabric -- empowering responsive and sustainable next-generation wireless. To enable decision-making in real time, Tiny \ac{ML} / Real-time \ac{ML} have emerged as key enablers, offering lightweight, responsive, and energy-efficient intelligence directly on wireless devices as illustrated in \fref{fig:connectedAI}. 

Tiny \ac{ML} focuses on deploying compact, power-efficient ML models on resource-constrained edge nodes such as sensors, controllers, and XR wearables. This makes it ideally suited for industrial automation and extended reality (XR) applications, where responsiveness, reliability, and energy efficiency are critical. For example, vibration sensors embedded with quantized convolutional neural networks (CNNs) can perform real-time anomaly detection, transmitting only significant events instead of raw data, which reduces both uplink load and radio energy consumption.

Let us consider an illustrative comparison between centralized GPU-based inference and edge-level TinyML deployment for a compact convolutional neural network performing object detection. A data-center GPU with a power envelope of approximately 250 W can execute inference in 5 ms, resulting in an energy consumption of roughly 1.25 J per inference. In contrast, for example, a TinyML model with 7 million parameters implementation on a microcontroller operating at 5-6 mW may require 300 ms per inference, yielding an energy expenditure of approximately 5 mJ. Although the TinyML solution incurs a tenfold increase in latency, it achieves more than two orders of magnitude lower energy consumption per task. This demonstrates a fundamental energy--latency trade-off: GPU platforms optimize throughput and computational intensity, whereas TinyML architectures prioritize energy efficiency and scalability for resource-constrained, battery-powered edge environments.

\subsection{Learning in real-time}
Energy-efficient real-time learning for connectivity and XR applications requires lightweight architectures combined with minimal training and inference scheduling using smaller-sized models. Such a schedule for real-time \ac{AI} systems should rely predominantly on fast adaptation to changing wireless environments by using online learning and statistical signal processing techniques. For example, this is essential in industrial scenarios where wireless channels are affected by mobility, reflections, and interference (e.g., on the production Shopfloor, adaptive beamforming using real-time \ac{AI} ensures stable and low-latency link updates for wireless motion control systems, even in highly reflective factory environments).

Fast-timescale updates (per signaling slot of 1 ms or per OFDM symbol of 66.7-down-to-1.04 $\mu$s) must be handled by adaptive filters (e.g, recursive least squares (RLS), Kalman filter) with tens to hundreds of parameters, while distributed stochastic gradient descent can directly adapt on-device TinyML models with 7 to 10 million parameters.

For medium-timescale schedule (i.e., minutes to hours), federated averaging of TinyML models ($\leq$ 10 M parameters) across connectivity or XR devices in Fig. \ref{fig:poc} is feasible.

In contrast, a large-timescale schedule of retraining in the edge cloud is infrequent (daily or event-driven) and limited to compressed large models (50-100 million parameters for control and up to 0.5-2 billion parameters for XR orchestration), ensuring limited energy consumption and predictable latency in industrial deployments.

Mobile \ac{AI} applications particularly benefit from edge intelligence through predictive rendering, where on-device TinyML models anticipate demand or motion to enable on-demand connectivity with the edge computing, reducing latency and bandwidth demand. Similarly, pruned neural networks running on AR headsets can locally refine the \ac{SLAM} system, while reinforcement learning agents at access points adjust resource allocation and beam directions to maintain latency with mobility under 5 ms. Depending on the scale, model quantization and pruning should further ensure sub-ms latency for a single token inference and predictable energy consumption for training.

Finally, combining the distributed learning concept with Real-time ML and Tiny \ac{ML} allows devices to collaboratively train lightweight models without exchanging raw data, preserving privacy and enabling intelligent factory operations across distributed wireless nodes.

\subsection{Augmenting Physical Context with Network Intelligence}
For the interpretation of the network and customer environment in real-time, it is expected that connected intelligence will be augmented with contextual information. This is achieved with an environment-specific {adaptive control for data collection} adjustable to device capabilities. A distributed controller needs to be designed for real-time orchestration on devices, taking into consideration spatiotemporal data. Therefore, as discussed in \sref{sec:dataChallenges}, the data controller is a prerequisite for efficient optimization and self-configuration. Here, the \ac{IRF} controller integrates the application's cyber-physical interface with \ac{AI}-programmable communication interfaces.

In principle, the embedded \ac{AI} computing as discussed in \sref{sec:embeddedComputing}, is expected to unlock powerful \ac{XR} applications across a range of 6G industrial applications. Figure~\ref{fig:poc} illustrates how \ac{AI}-based \ac{XR} systems interact with the underlying connectivity nodes and local/cloud computing systems. The architecture is organized in two segments: the local \ac{XR} application computing, responsible for real-time \ac{XR} performance, and edge-based \ac{AI}-\ac{XR} computing, responsible for data-driven adaptation and optimization. At the foundation, as depicted in the figure, the node functions provide high-performance connectivity and feedback between \ac{XR} applications and the communication hardware. This modular structure enables real-time, bidirectional flows of data and control between the \ac{XR} application layer and the \ac{RAN}, making it possible for \ac{AI}-based \ac{XR} solutions to deliver interactive interfaces, predictive analytics, and reliable communication for mobile \ac{AI} applications.

\subsection{Data control in real time}\label{sec:dataChallenges}
In a virtual emergency application, mobile \ac{AI} applications require data from a wide variety of distributed sensors (e.g., health monitors, environmental detectors, and imaging systems) to be collected, processed, and acted upon in real time. However, the nature of this data is heterogeneous, both in structure and generation patterns. The system cannot rely on predefined volumes, locations, or temporal consistency in the data streams. This necessitates \ac{AI}-programmable control over the {distribution, frequency, and volume} of data collection, such that communication, computing, and sensing resources are efficiently utilized without service interruptions or unnecessary traffic overhead. For example, collecting too much data risks congesting the network and wasting energy, while insufficient or infrequent data compromises inference quality and undermines root-cause analysis. 
Centralized (i.e., big data) approach breaks down in terms of scalability and energy efficiency, especially when new nodes or sensing sources are dynamically added or removed in a spatiotemporal fashion, when both inference and training must occur in ultra-short signaling intervals (see \sref{sec:responsivenessChallenge}). 
This is because real-time decision-making demands immediate responses to new data based on application-specific and short-burst data streams, i.e., fine-tuned spatiotemporal data computing.

Moreover, in ultra-dense deployments, data often exhibits strong spatial correlation, where distributed sensors may capture different perspectives of the same communication event. The key challenge becomes not only pre-processing these overlapping data streams for efficient learning, but also ensuring that data volumes are sufficient and timely. Otherwise, delayed or sparse data collection may render the observations unusable.

\subsection{Relevance to Standardization}
Importantly, 3GPP specifications (3GPP TS 23.288; 3GPP TS 28.105; 3GPP TS 28.535; 3GPP TR 23.700) define intent-driven management, automated radio access optimization, and standardized training/inference coordination that is necessary for the design of AI-programmable wireless connectivity. Similarly, the Open RAN architecture (O-RAN Alliance) defines functional disaggregation (O-RU, O-DU, O-CU), open fronthaul, and AI-enabled control via near-Real-Time (RT) and non-RT RAN Intelligent Controllers (RICs, O-RAN WG1/WG2). Non-RT RICs support rApps for long-term model training and policy optimization, while near-RT RICs host xApps for 10 ms up to 1 s control loops via the E2 interface (radio access, mobility, and energy management). The IRF vision in this paper extends these contributions by integrating radio and edge resources into a unified AI-programmable connectivity with energy-efficient computing for latency-constrained immersive XR applications.

\section{Spatial Computing in Communications}\label{sec:dev}
Next, an example is conceptualized where \ac{XR}-based spatial computing is integrated with a distributed, overlay, and asynchronous device communication network for field technician support since the impact of \ac{AI} models and \ac{XR} technologies on the telecom industry is expected to be significant in customer services and networking areas \cite{wwf}. 
For example, the ever-increasing complexity of private networks introduces new challenges for industrial field technicians responsible for installation, troubleshooting, and maintenance. Spatial computing, combined with \ac{AI} and \ac{XR}, provides a novel paradigm to enhance technician efficiency and ensure resilient industrial connectivity.

\subsection{Field-technician Example}
Consider the case of a digital factory equipped with a private network. When a connectivity fault occurs, such as reduced coverage or interference, field technicians can leverage a spatially aware digital twin of the factory floor. The digital twin integrates real-time sensor data, wireless performance metrics, and moving robot (drone) telemetry, allowing technicians to visualize spatiotemporal network behavior through \ac{XR} devices \cite{shafi2025}. For instance, an AI-enabled XR headset may project a coverage heatmap directly onto the production environment, highlighting problem zones caused by antenna misalignment or signal obstruction from newly installed machinery.

Instead of relying on traditional trial-and-error adjustments, the technician can interact with the digital twin to simulate corrective actions, such as antenna tilt modifications, beamforming configurations, or frequency adjustments. The system evaluates and visualizes predicted outcomes before changes are applied to the physical infrastructure, thereby minimizing downtime and operational disruption. In cases requiring additional expertise, the technician can share the immersive digital environment with remote experts, who can “step into” the virtual representation of the factory to collaboratively analyze interference patterns and propose solutions. Over time, operational data gathered through spatial computing contributes to predictive maintenance models, enabling proactive interventions and strengthening the resilience of private networks in digital factories.

\begin{figure}[t]
\centerline{\includegraphics[width=\linewidth]{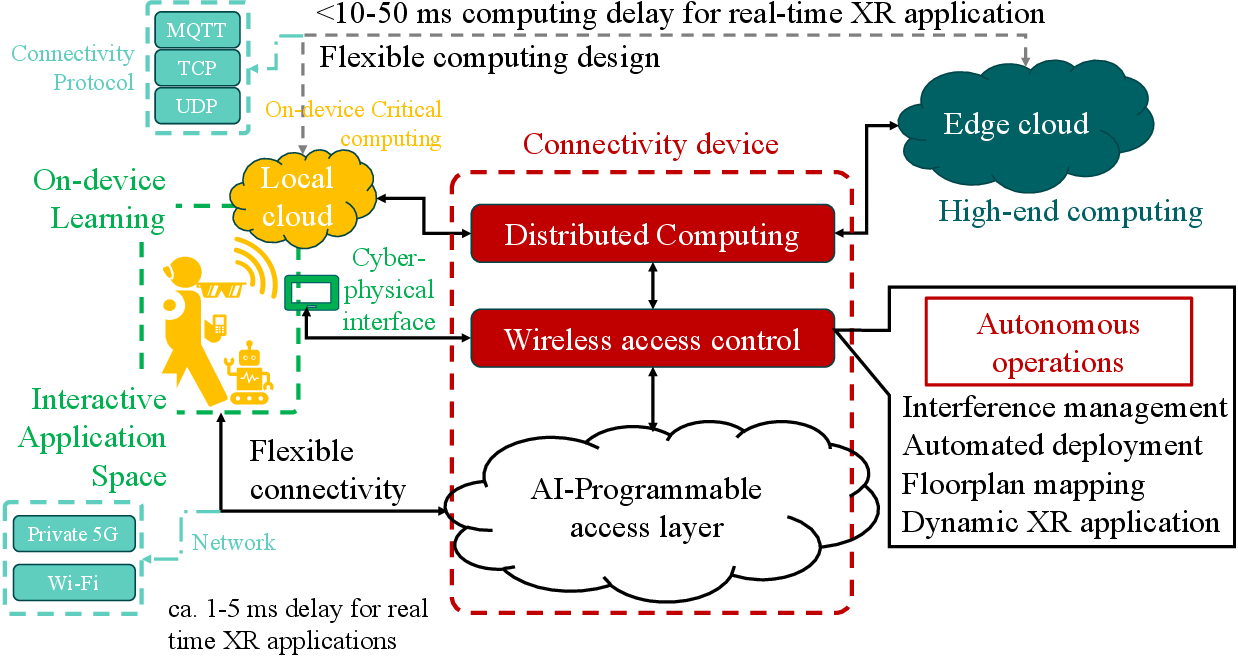}}
\caption{6G device concept of integrated mobile \ac{AI} and \ac{XR} applications.}
\label{fig:poc}
\end{figure}

\subsection{Spatial Computing and Communication Architecture}
The illustration in \fref{fig:poc} conceptualizes the \ac{IRF} architecture integrating spatial computing and communication functions, including on-device local- and edge-cloud computing, spatial computing, and distributed learning. 

In the flexible federated/distributed XR architecture shown in \fref{fig:poc}, data throughput and computing resources are dynamically split among the XR device, the local cloud, and the edge cloud, according to latency and energy constraints. 
For a truly immersive XR experience, data throughput is characterized by periodic, high-volume bursts (e.g., 120 frames per second with resolutions up to 8 K). Therefore, with split rendering, immersive \ac{XR} requires tens of Mbps radio access with a latency in the order of 10 ms measured at the transport/application layer \cite{3gpp-xr}.

Ultra-latency-critical tasks such as sensor fusion, pose estimation, eye tracking, and prediction are executed locally on the device ($\approx$10--50 Trillions of Operations Per Second (TOPS)) to maintain sub-10 ms motion-to-photon latency according to 3GPP TR 26.998 specification. For example, we expect a 3 ms latency by a tethered ROG XREAL R1 device (to be released in 2026) with all computing done on the computer, while the display function is operated in the glasses. On the other hand, current devices with on-device processing from Meta and HTC have Motion-to-photon latency in VR (AR) mode of about 2--3 (11--40) ms, while Apple Vision Pro offers 11 ms AR mode latency. Specifically, local models are continuously updated (tens to hundreds of MB per single round) and optimized based on channel state, available bandwidth, battery level, network congestion, and device thermal constraints.

Pre-computed data and compressed foveated video streams (typically 100-600 Mbps) are exchanged between the local or edge cloud in \fref{fig:poc}, where compute-intensive functions such as dense Simultaneous Localization and Mapping (SLAM) back-end processing, multi-user scene fusion, experience emulation, and large-model inference are performed using GPU clusters (e.g., 10 Tera Floating-Point Operations Per Second (TFLOPS) to Peta Floating-Point Operations Per Second (PFLOPS) scale). Accordingly, the local cloud enables deterministic low-latency XR applications having the real world as the zero-latency reference, while the network-supported edge provides scalable rendering and foundation-model support.

The main functions are described below: 

\textbf{[Distributed Computing]} Signal processing and \ac{AI} methods are deployed to design an efficient communication and computing architecture as proposed in \cite{ieice_native_ai_dsp}. The integration of learning with the low-latency and high-reliability requirements is achieved by implementing lightweight learning models. Such mechanisms enable the adaptive optimization of transceiver functions, including dynamic beamforming adjustments in response to user mobility, fast interference suppression in dense networks, and the recognition of real-time modulation and coding schemes under varying channel conditions. 

Distributed computing is, for example, implemented by a Leader Election algorithm designed for improving the coordination of the infrastructure transceivers for a distributed or centralized training location. A single process is defined as the organizer of training locally or distributed between several transceivers. On the other hand, wireless access control determines how to schedule transceivers to locally explore their environment while minimizing the impact on each other and stabilizing overall learning. To decrease the mean time to failure of transceivers, we aim to increase the number of computing and communications devices. Thus, a locally controlled message-passing interface is defined for communication to share the data from the locally adjacent devices to facilitate on-device embedded learning.

\textbf{[On-device learning]} The communication architecture supports active learning, split between local-cloud and edge-cloud, with integrated knowledge management \cite{ai4coms, ieice_native_ai_dsp}, while considering the latency and reliability constraints of \ac{XR} through AI-programmable access. In this example, distributed local- and edge-based training of the learning agent (i.e., reinforcement learning or deep learning) can be implemented as distributed Q-learning, policy-gradient, or the actor-critic deep deterministic policy gradient algorithm \cite{self-opt}.

In synchronous distributed training with GPUs (i.e., workers) in ultra-dense communication systems, the iteration time is dominated by the slowest GPU (i.e., the ``straggler''), since the parameter device/server must wait for all gradients before updating the model. This leads to poor scalability as the number of GPUs increases. On the other hand, when federated or distributed learning is invoked, the synchronization strategy directly impacts adaptability and reliability. For sub-$\mu$s decision-making aligned with a 4.17~$\mu$s OFDM symbol duration (e.g., 240~kHz subcarrier spacing), fully round-based synchronization may introduce excessive latency due to barrier waiting. Instead, semi-asynchronous aggregation with bounded staleness enables continuous model refinement while respecting tight scheduling constraints (e.g., maximum delay $\tau \leq$ 2--3 local updates or $\leq$ 50~$\mu$s clock skew between two different GPU clocks). Here, the slowest GPUs are handled by deadline-based aggregation (e.g., 90\% client participation within a 100~$\mu$s aggregation window) and/or weighted updates to compensate for partial participation.

To maintain symmetric Gbps throughput (e.g., uplink/downlink of 0.2-1 Gbps), the communication overhead must remain below 1-3\% of radio resources. To reduce control-plane throughput to the Mbps range per computing device, gradient compression techniques can be applied, including quantization (i.e., 8-bit or 4-bit quantization reduces the numerical precision of each gradient value exchanged between \ac{GPU}s), sparsification by transmitting only the most relevant gradients (i.e., Top-$k$ selection with $k=1-5\%$ of gradients with the largest absolute magnitudes exchanged between \ac{GPU}s), and periodic model averaging (\ac{GPU}s train on their local data independently for 10-20 local steps before synchronizing and averaging their model weights). These techniques reduce the control-plane load to sub-10 Mbps per node, ensuring reliable coverage while preserving adaptability in distributed learning-enabled wireless systems.

The AI-programmable access framework enables flexible connectivity and intelligent wireless access control, supporting both 3GPP-based private networks and IEEE 802.11 random access systems \cite{ieice_native_ai_dsp}. To realize this, an AI-programmable \ac{API} is defined to facilitate automation and fine-grained control of transceiver functions (see \tref{tab:api}). The \ac{API} also extends to AI- and XR-enabled application devices, leveraging \ac{M2M} communication via the \ac{MQTT} protocol, while integrating local and edge-cloud computing resources for low-latency processing and coordination as illustrated in \fref{fig:poc}.

\textbf{[Interative Application]} For autonomous operation, for example, a spatial \ac{XR}-computing is introduced to track and interactively map Tx/Rx beam locations, while mapping and analyzing the overall system performance by using 3D heatmaps, spatial Anchors, and sophisticated graphs and charts on 3-dimensional dashboards to display simplified high-level data insights \cite{shafi2025}. 

When defining the \ac{API}s, one must consider the latency and computational constraints of the wireless system, as well as the hardware limitations of connected devices, such as memory space and computational power. The \ac{API}s need to be implemented on an open network platform based on \ac{SDR}.

\section{Conclusions}
This vision paper discussed the communications, computing, and sensing requirements of future interactive applications. The major requirements are discussed around the three aspects: (1) Online training operation in real-time with ultra-low energy consumption, (2) Real-time data sensing at the speed of the target radio application, and (3) Sub-ms decision-making and efficient on-device processing. The research and development limitations of enabling communication technologies are discussed, such as distributed connectivity, embedded learning, and spatial \ac{XR} computing to meet the technological demands of interactive applications. Ultimately, achieving energy-efficient, real-time learning and wireless sensing requires a paradigm shift in how intelligence is embedded and coordinated across the wireless devices. This includes the co-design of communication protocols and hardware interfaces to support distributed and adaptive learning under strict energy, latency, and computational constraints.

It should be noted that distributed AI within ultra-dense IRF may pose risks to privacy, data integrity, and reliability (e.g., gradient leakage, falsified updates, manipulated sensor data, adversarial attacks, congestion, and model divergence), and privacy-preserving, secure, and resilient mechanisms are essential to ensure safe, reliable operation of \ac{IRF}.


\section*{Acknowledgement}
This work was partially funded by the German Federal Ministry of Research, Technology, and Space (BMFTR) within the 6GEM+ Transfer Hub, GEM-X, and VICTOR6G projects, respectively, under the grant numbers 16KIS2409K, 16KISS004K, and 16KIS2547.

\bibliographystyle{IEEEtran}
\bibliography{references}

@ARTICLE{tinyML1,
  author={Wu, Guanghan and Tarkoma, Sasu and Morabito, Roberto},
  journal={IEEE Internet of Things Magazine}, 
  title={Consolidating TinyML Lifecycle With Large Language Models: Reality, Illusion, or Opportunity?}, 
  year={2025},
  volume={8},
  number={5},
  pages={88-96},
  doi={10.1109/MIOT.2025.3575927}}

@INPROCEEDINGS{tinyML2,
  author={El Zeinaty, Christophe and Hamidouche, Wassim and Herrou, Glenn and Menard, Daniel and Debbah, Merouane},
  booktitle={2025 International Joint Conference on Neural Networks (IJCNN)}, 
  title={Can LLMs Revolutionize the Design of Explainable and Efficient TinyML Models?}, 
  year={2025},
  volume={},
  number={},
  pages={1-10},
  doi={10.1109/IJCNN64981.2025.11229057}}

@ARTICLE{tinyML3,
  author={López, Onel and Singh, Ritesh Kumar and Phan-Huy, Dinh-Thuy and Katranaras, Efstathios and Mazloum, Nafiseh and Ruttik, Kalle and Jäntti, Riku and Khan, Hamza and Rosabal, Osmel and Alexias, Pavlos and Raghuwanshi, Prasoon and Ruiz-Guirola, David and Singh, Bikramjit and Höglund, Andreas and Van, Dung Pham and Azarbahram, Amirhossein and Famaey, Jeroen},
  journal={IEEE Internet of Things Magazine}, 
  title={Zero-Energy Devices for 6G: Technical Enablers at a Glance}, 
  year={2025},
  volume={8},
  number={3},
  pages={14-22},
  doi={10.1109/IOTM.001.2400138}}

@INPROCEEDINGS{agenticAI,
  author={Tong, Wen and Huo, Wei and Lejkin, Thierry and Penhoat, Joel and Peng, Chenghui and Pereira, Carlos and Wang, Fei and Wu, Shaoyun and Yang, Lu and Shi, Yuanming},
  booktitle={2025 IEEE International Conference on Communications Workshops (ICC Workshops)}, 
  title={A-Core: A Novel Framework of Agentic AI in the 6G Core Network}, 
  year={2025},
  volume={},
  number={},
  pages={1104-1109},
  doi={10.1109/ICCWorkshops67674.2025.11162291}}

@ARTICLE{holocomms,
  author={K. Wakunami and P.-Y. Hsieh and R. Oi and T. Senoh and H. Sasaki and Y. Ichihashi and M. Okui and Y. Huang and K. Yamamoto},
  journal={Nature Communications}, 
  title={Projection-type see-through holographic three-dimensional display}, 
  year={2016},
  volume={},
  number={},
  pages={},
  doi={}}

@ARTICLE{ieice_native_ai_dsp,
  author={F. B. Saghezchi and M. Pourghasemian and B. Ding and A. Abdi and B. Lee and A. Baron and M. Shafi and H. Gacanin},
  journal={IEICE Trans. on Communications, Special Section on Smart Radio Technologies}, 
  title={AI-Native Radio Transceiver Signal Processing for Next-Generation Mobile Communication Systems}, 
  year={2025},
  volume={September},
  number={},
  pages={},
  doi={}}

@ARTICLE{ai4coms,
  author={H. Gacanin},
  journal={IEEE Vehicular Technology Magazine, Special issue on 6G: What is Next?}, 
  title={Autonomous Wireless Systems with Artificial Intelligence: A Knowledge Management Perspective}, 
  year={2019},
  volume={September},
  number={},
  pages={51 - 59},
  doi={}}

@unpublished{shafi2025,
  author       = {M. Shafi and M. Canoglu and H. Guang and H. Gacanin and K. Kondo and Y. Shoji},
  title        = {{Real-time XR Visualization for Terahertz Transceiver System}},
  note         = {{Germany-Japan Beyond 5G/6G Workshop}, Tokyo},
  year         = {January 2025}
}

@ARTICLE{self-opt,
  author={H. Gacanin and E. Perenda and S. Karunarathne and R. Atawia},
  journal={IEEE Trans. on Vehicular Technology}, 
  title={Self-optimization of Wireless Systems with Knowledge Management: An Artificial Intelligence Approach}, 
  year={2019},
  volume={68},
  number={10},
  pages={9682-9697},
  doi={}}

@ARTICLE{6g-ai-standard,
  author={S. Majumdar and Q. Wei and S. Schwarzmann and R. Trivisonno and G. Carle},
  journal={IEEE Communications Standards Magazine}, 
  title={Towards AI-Native 6G Systems: Standards Enablers for 6G Network Automation}, 
  year={2015},
  volume={},
  number={},
  pages={},
  doi={}}

@techreport{ericsson-ainative,
  title        = {AI-Native: Making Intelligent Networks a Reality},
  author       = {Ericsson},
  institution  = {Ericsson},
  year         = {2024},
  type         = {White Paper},
  url          = {https://www.ericsson.com/en/reports-and-papers/white-papers/ai-native},
  note         = {Accessed: 2025-05-21}
}

@techreport{3gpp-xr,
  title        = {Architecture Enhancements for XR (Extended Reality) and media service},
  author       = {3GPP Working Group SA2, Study on XR (Extended Reality) and media services - Release 18},
  institution  = {3GPP},
  year         = {2022},
  type         = {Standard specification TR 23.700-60},
  url          = {},
  note         = {Accessed: 2026-01-18}
}

@techreport{3gpp-ai,
  title        = {Artificial Intelligence/Machine Learning (AI/ML) management},
  author       = {{Technical Specification Group Services and System Aspects Management and Orchestration - Release 19}},
  institution  = {3GPP},
  year         = {2025},
  type         = {Standard specification},
  url          = {},
  note         = {Accessed: 2025-04-18}
}

@techreport{wwf,
  title        = {Quantifying the Mutual Requirements
Driving AI and 6G Co-Evolution},
  author       = {L. Bariah and B. Mefgouda and A. D. Domenico and L. Bariah and C. Marquezan and R. Trivisonno and Merouane Debbah},
  institution  = {Wireless World Research Forum},
  year         = {2025},
  type         = {White Paper},
  url          = {https://wwrf.ch/publications/outlook-publications/},
  note         = {Accessed: 2025-06-25}
}

@techreport{nokia2024ai,
  title        = {{AI for Radio Access Networks}},
  author       = {{Nokia}},
  institution  = {Nokia Corporation},
  year         = {2024},
  type         = {White Paper},
  url          = {https://www.nokia.com/networks/white-papers/ai-for-radio-access-networks/},
  note         = {Accessed: 2025-05-21}
}

%
\begin{IEEEbiography}[{\includegraphics[width=1in,height=1.5in,clip,keepaspectratio]{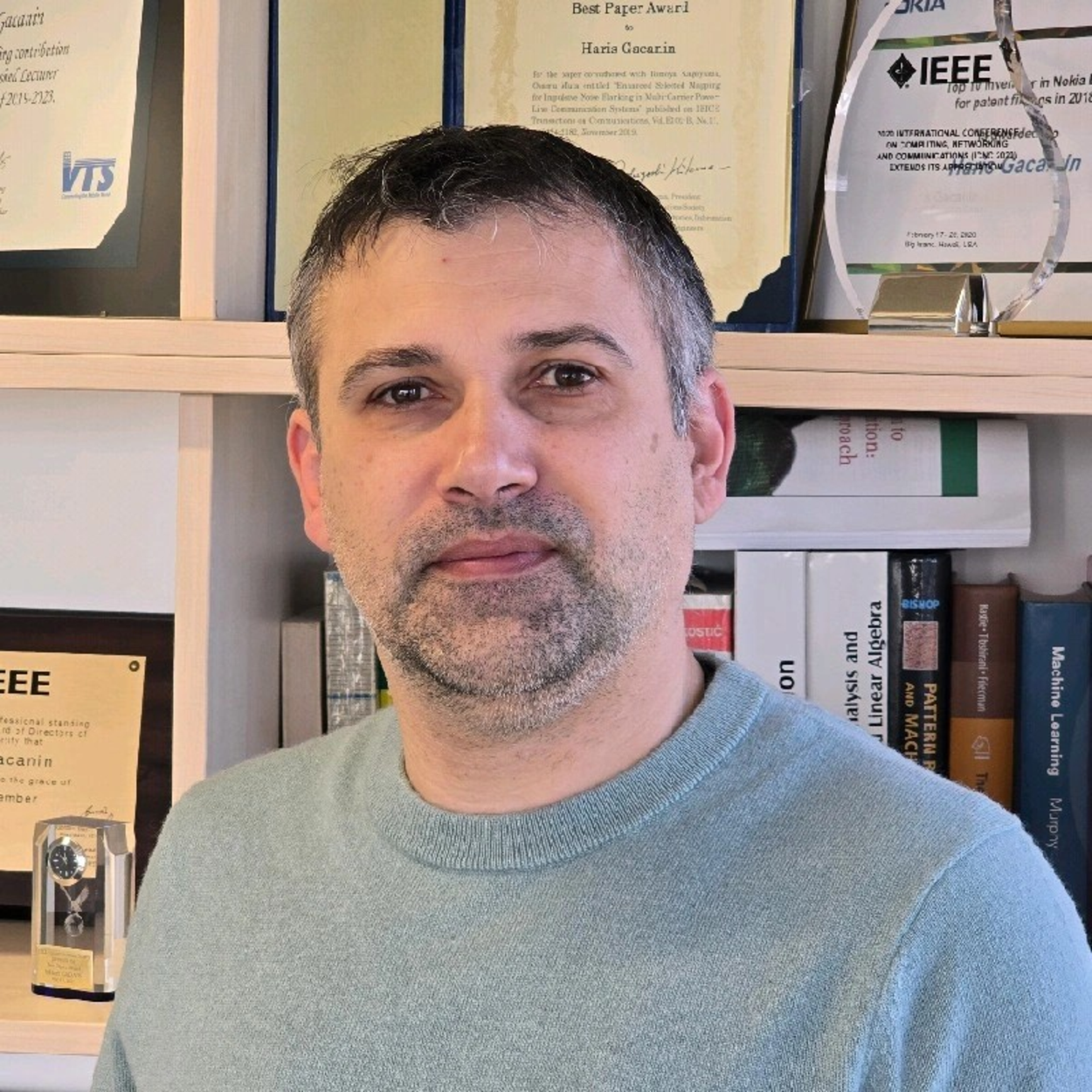}}]{Haris Gačanin} (IEEE Fellow’21) received the Ph.D. degree from Tohoku University, Japan, in 2008. He was an Assistant Professor with Tohoku University before joining Alcatel-Lucent Bell (now Nokia Bell Labs) in 2010, where he served as a Physical-Layer Expert and later as Department Head. Since April 2020, he has been a Full Professor at the Chair for Distributed Signal Processing at RWTH Aachen University in Aachen, Germany. His research interests include broad areas of signal processing and machine learning in wireless communications.
\end{IEEEbiography}

\end{document}